\documentclass[final,5p,times,twocolumn]{elsarticle} 

\usepackage{graphicx}

\usepackage{float}
\usepackage{gensymb}

\usepackage{amssymb}

\usepackage{lineno}
\usepackage{subcaption}

\begin{document}

\begin{frontmatter}



\title{A Monolithic active pixel sensor for ionizing radiation using a 180nm HV-SOI process} 



\author{Tomasz Hemperek}
\ead{hemperek@uni-bonn.de} 


\author{Tetsuichi Kishishita}
\author{Hans Kr\"uger}
\author{Norbert Wermes}

\address{Physikalisches Institut, Universit\"at Bonn, Nussallee 12, 53115 Bonn, Germany} 


\begin{abstract}
An improved SOI-MAPS (Silicon On Insulator Monolithic Active Pixel Sensor) for ionizing radiation based on thick-film High Voltage SOI technology (HV-SOI) has been developed. Similar to existing Fully Depleted SOI-based (FD-SOI) MAPS, a buried silicon oxide inter-dielectric (BOX) layer is used to separate the CMOS electronics from the handle wafer which is used as a depleted charge collection layer. FD-SOI MAPS suffer from radiation damage such as transistor threshold voltage shifts due to charge traps in the oxide layers and charge states created at the silicon oxide boundaries (back gate effect). The X-FAB 180-nm HV-SOI technology offers an additional isolation by deep non-depleted implant between the BOX layer and the active circuitry witch mitigates this problem. Therefore we see in this technology a high potential to implement radiation-tolerant MAPS with fast charge collection property. The design and measurement results from a first prototype are presented including charge collection in neutron irradiated samples.
\end{abstract}


\end{frontmatter}

\section{Introduction}
\label{}
Monolithic Active Pixel based on SOI technology has been proposed as an ultimate monolithic sensor approach for tracking detectors due to the fact that the sensor and front-end readout electronics with different requirements on silicon parameters can be integrated into a single chip \cite{Amati}.
Such a technology offers fabrication of devices with a large number of readout channels with fine segmentation at a small cost.

The first prototypes of an SOI-based MAPS were implemented with a FD-SOI technology \cite{Bulgheroni}, in which the whole body under the transistor gate is completely depleted. Unfortunately, the FD-SOI process by principle suffers significantly from Total Ionizing Dose (TID) effects \cite{Kochiyama}. Several ways have been investigated to mitigate this problem, such as extra isolating implants in the handling wafer, or the use of double SOI wafers\cite{Honda}.

A new commercial HV-SOI process \cite{Holke}, that mitigates the back gate effect problem, is being investigated. This process offers n- and p-well structures for active layers, which creates an additional isolation layer between BOX and the active circuitry making transistor parameters insensitive to radiation effects in the BOX. 

\begin{figure}
\centering
\begin{subfigure}[b]{0.4\textwidth}
  \centering
  \includegraphics[width=1\linewidth]{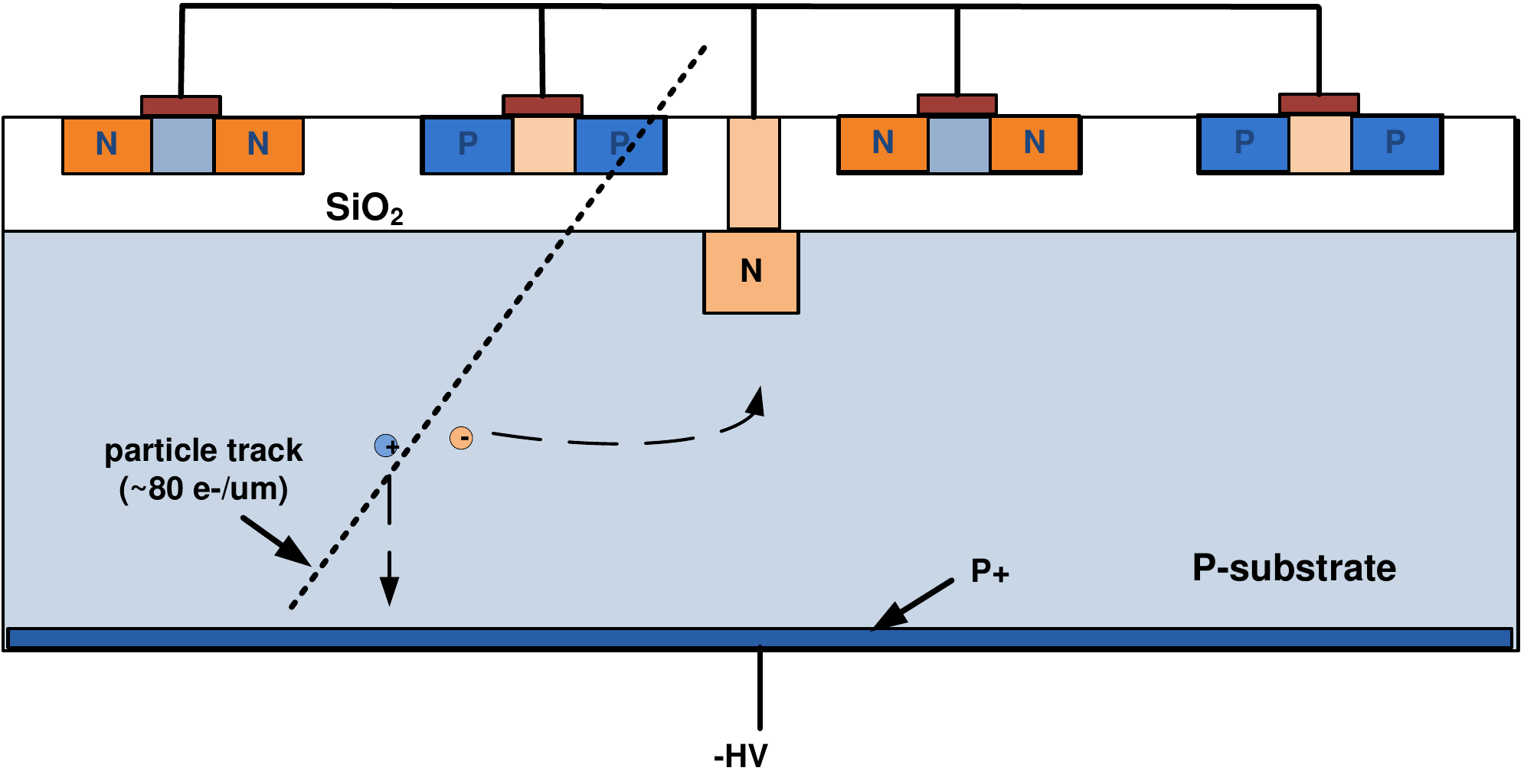}
  \caption{}
  \label{fig:hv_soi_cross}
\end{subfigure}%

\begin{subfigure}[b]{0.4\textwidth}
  \centering
  \includegraphics[width=1\linewidth]{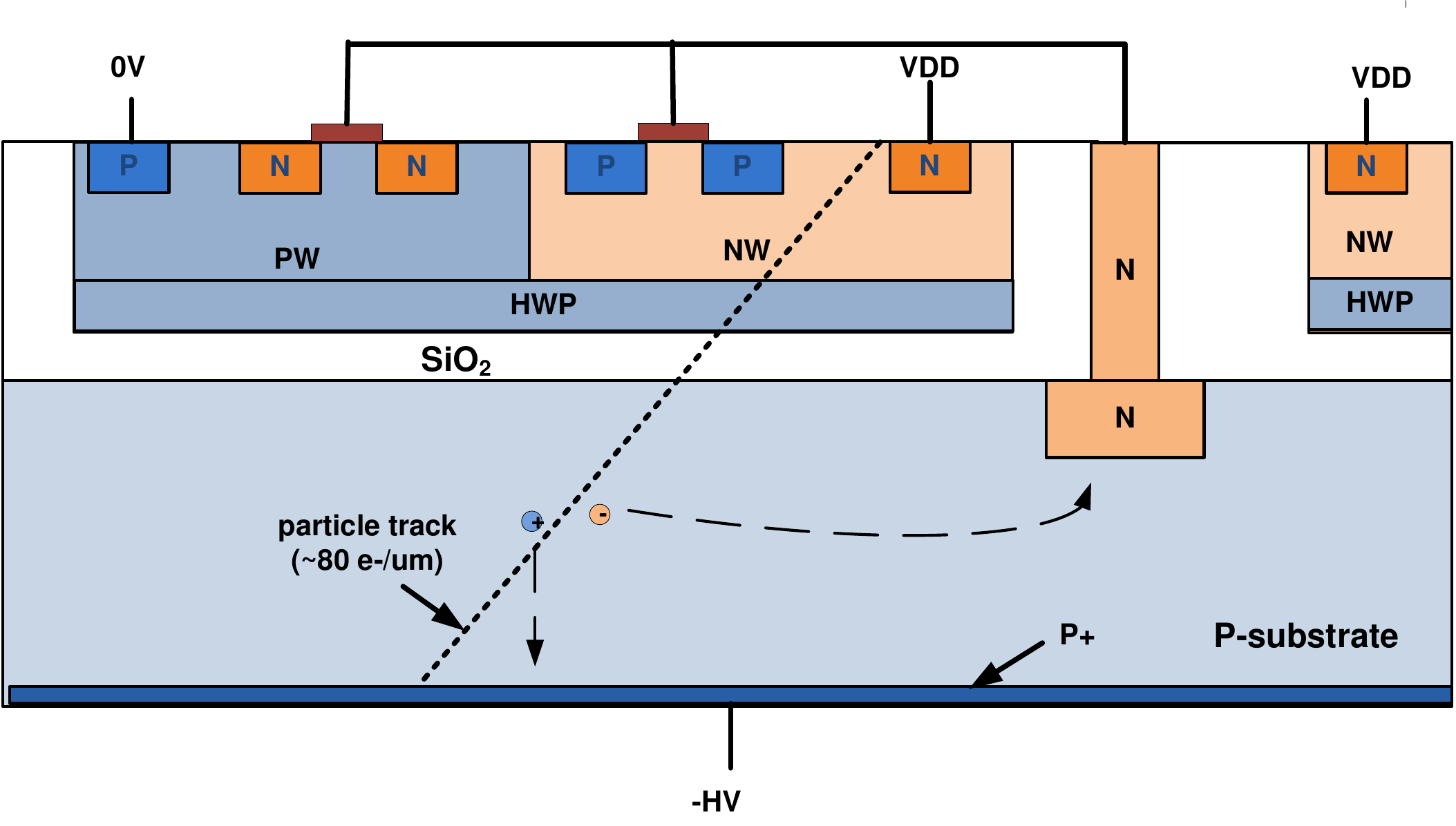}
  \caption{}
  \label{fig:fd_soi_cross}
\end{subfigure}
\caption{Cross section of  a) the FD-SOI and b) the thick-film HV-SOI process for p-type silicon substrate and n-type collection diode}
\label{fig:fd_pd_soi_cross}
\end{figure}

To evaluate the process, we designed the prototype chip ``XTB01'', which includes simple diode and readout electronics. In this article, the concept and measurement results of the prototype chip implemented with this technology are presented. The TID effects on this HV-SOI technology is reported in a separate publication \cite{Sonia}.  

\section{Technology Overview}

A simplified cross-section of FD-SOI and thick-film SOI technologies is shown in Figure~\ref{fig:fd_pd_soi_cross}. In a SOI process, active devices are fabricated in a thin silicon layer on top of an insulating layer of silicon dioxide (buried oxide). The inactive layer underneath BOX (handling wafer) can be used as a depleted sensor layer. The main difference of thick-film SOI technology is that the active (transistor) layer is thick (few~$\mu$m) compared to tens of nm in case of FD-SOI. Multiple biased well structures give the possibility to isolate the transistor from any influence of charge build-up in the BOX. Both technologies allow access to the handling wafer to create a charge collecting diode and bias. A standard CMOS circuit can be realized in the active layer. The process itself gives us a possibility to create high voltage (above 200V) transistors for power application. The expected depletion thickness of the handling wafer in this prototype is about 50~$\mu$m. 
Table \ref{table:tech_potions} summarizes the technological overview. 

\begin{table}[h!]
  \begin{center}
    \caption{Overview of technological options for the prototype chip.}
    \label{table:tech_potions}
    \begin{tabular}{ l | c }
    \hline \hline
    Features & 180 nm, 4 metal layers, SOI  \\
    Supply rail& 1.8V  \\
    Handling wafer type & p-type bulk with 100~$\Omega\cdot$cm \\
    Process options & MIM capacitor, deep HV well option  \\
    Chip area & 2x5 mm$^2$  \\
    \hline \hline
    \end{tabular}
  \end{center}
\end{table}
\begin{figure}[H]

\centering
\begin{subfigure}[b]{.5\textwidth}
  \centering
  \includegraphics[width=1\linewidth]{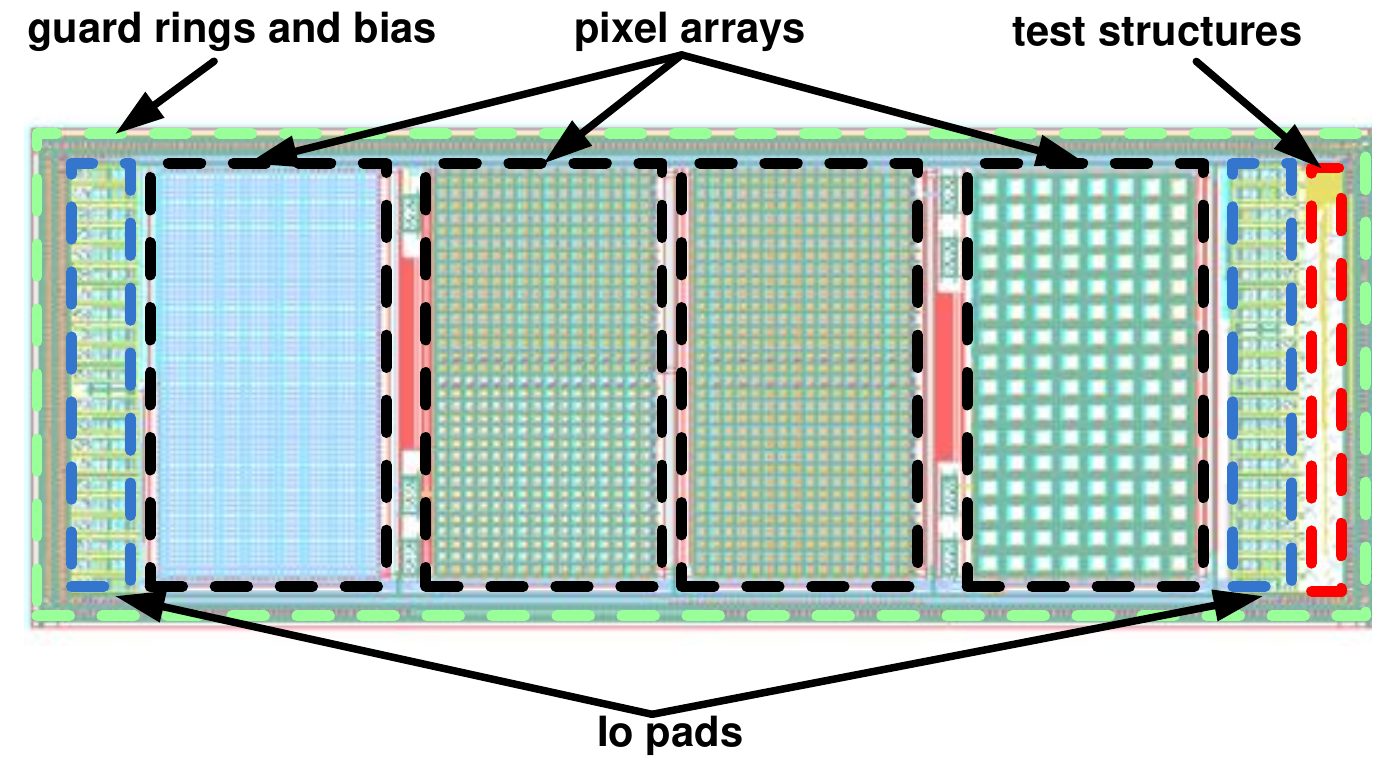}
  \caption{}
\end{subfigure}%

\begin{subfigure}[b]{.5\textwidth}
  \centering
  \includegraphics[width=1\linewidth]{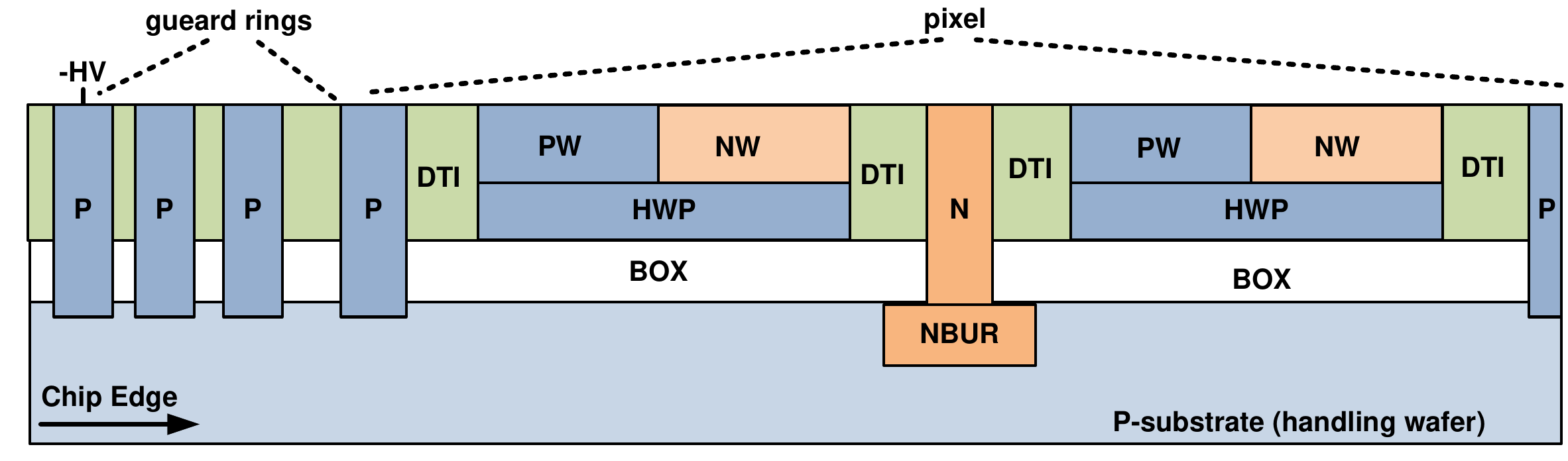}
  \caption{}
\end{subfigure}
\caption{XTB01 prototype a) layout  b) cross section}
\label{fig:xtb01_picture_cross}
\end{figure}

\section{Test Chip XTB01}

To investigate the feasibility of the particle detector based on this HV-SOI technology a prototype chip XTB01 has been designed with a simple diode structure and readout scheme.
The device consists of 4 pixel arrays with three different pixel sizes (25, 50, 100~$\mu$m). 
For the time being, no back implant is being used and thus HV bias is applied laterally. The HV ring surrounds every pixel and a multi-guard ring structure is placed next to the chip edge.
A picture and cross section of the chip can be seen in Figure~\ref{fig:xtb01_picture_cross}.
We also implemented an array of standalone P and N type transistors, i.e., without any sensor diodes connected, at the periphery of the chip, to check immunity against radiation effects.

\subsection{Pixel Design}

\begin{figure}
  \centering
  \includegraphics[width=1\linewidth]{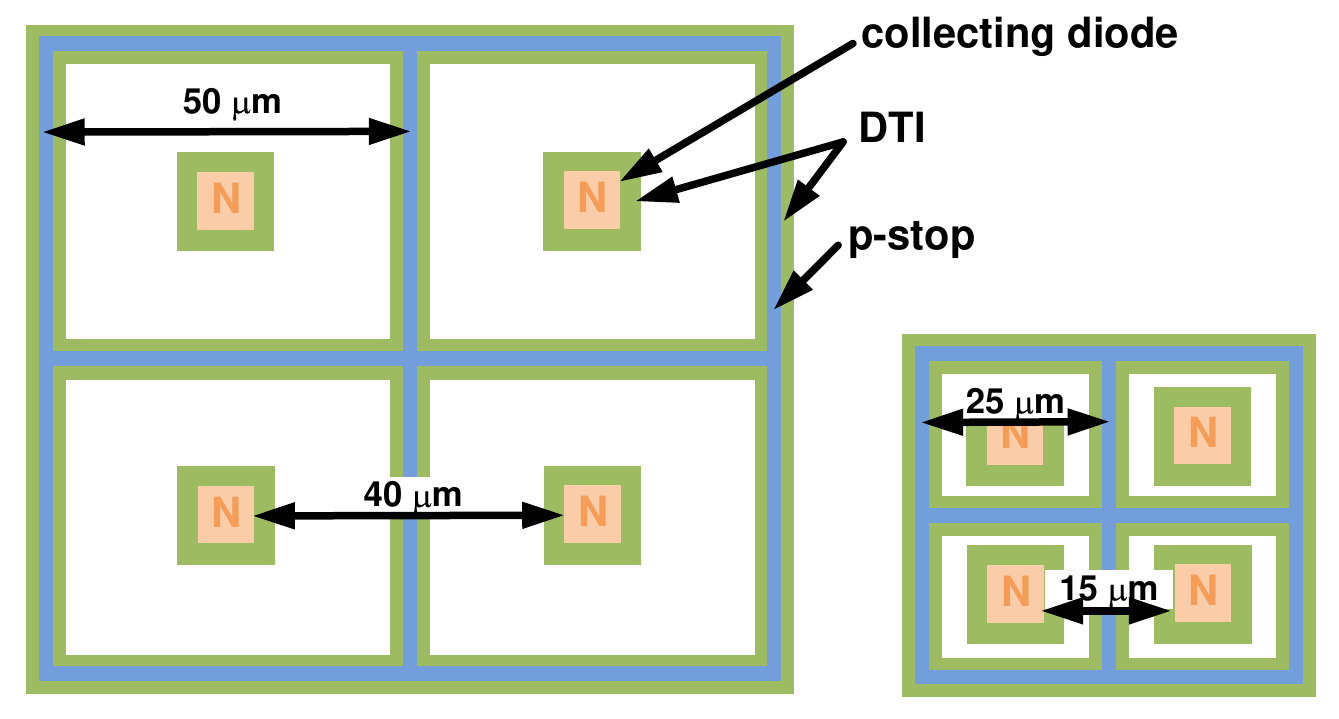}
  \caption{Pixel layout for 50~$\mu$m and 25~$\mu$m pitch pixels}
  \label{fig:pix_box_layout}
\end{figure}

Since the handling wafer material is p-type, n-type implants are used as a collecting electrode.
The readout transistors are placed adjacent to the collecting electrode. This region is separated from the diode by deep trench isolation (DTI). The implant structures for 25 and 50~$\mu$m pixels are shown in Figure~\ref{fig:pix_box_layout}. P-type (p-stop) openings are placed between pixels to "break" the electron accumulation layer underneath the BOX in particular after radiation.

\begin{figure}[H]
\centering
\includegraphics[width=0.5\textwidth]{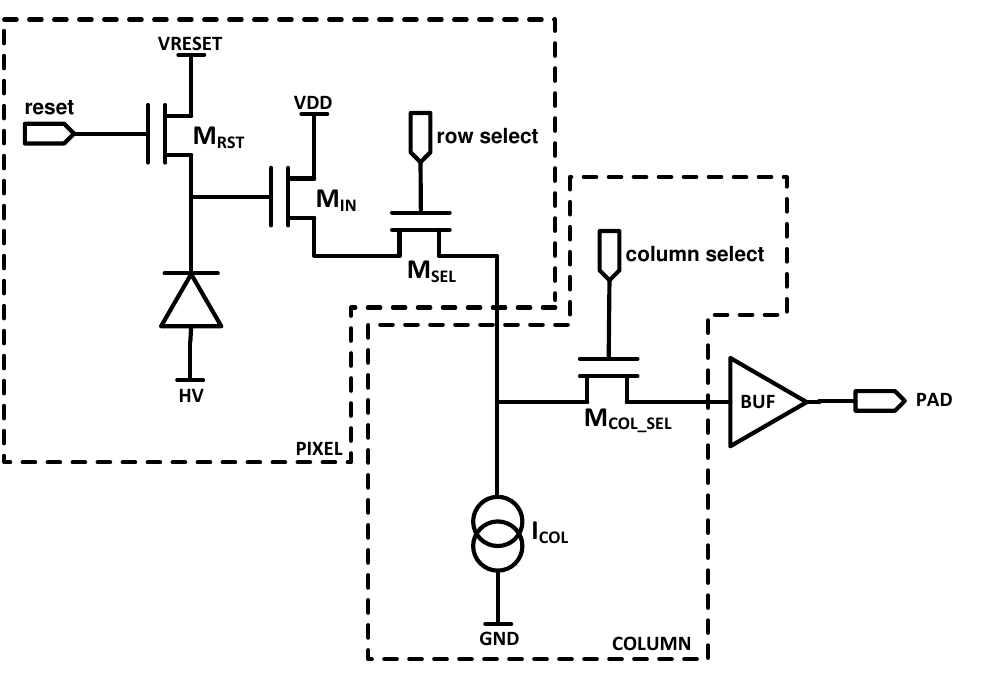}
\caption{Block diagram of 3T pixel readout for XTB01 chip.}
\label{fig:readout}
\end{figure}

\subsection{Readout Scheme}

The majority of pixels are standard three transistor (3T) pixels that allow direct access to the analog signal. 
This 3T readout scheme is widely used in various CMOS image sensors, and it is easy to compare sensor characteristics, such as the diode leakage current or noise performance, with other technologies. Figure~\ref{fig:readout} shows a block diagram of the 3T readout scheme \cite{Turchetta}. The reset transistor $M_{\rm RST}$ is used to reset the pixel by dumping the integrated charge to the positive bias voltage. The transistor $M_{\rm SEL}$ is activated to select the readout of the pixel, and $M_{\rm IN}$ is the input transistor of a source follower. The current source is common to all the pixels in one column. A signal integrated on input capacitance is directly proportional to the charge collected and scaled by input capacitance. The control signals are provided by two shift register arrays for row and column selection. One pixel is read one at a time in rolling-shutter manner. To investigate the optimum size of the input transistors against the diode size, we added variations on transistor widths and geometries.

\begin{table}[h!]
  \begin{center}
    \caption{Irradiation levels at different steps. (TID dose from reactor background.)}
    \label{table:irrad}
    \begin{tabular}{ c | c }
    \hline \hline
    Fluence [ n$_{\rm eq}$/cm$^{2}$] & Dose [kRad]\\
    \hline
    0 & 0 \\
    $1\times10^{13}$ & 10  \\
    $5\times10^{13}$ & 50  \\
    $1\times10^{14}$ & 100 \\
    $5\times10^{14}$ & 500 \\
    \hline \hline
    \end{tabular}
  \end{center}
\end{table}

\section{Sensor performance}

\begin{figure}
\centering
\includegraphics[width=0.5\textwidth]{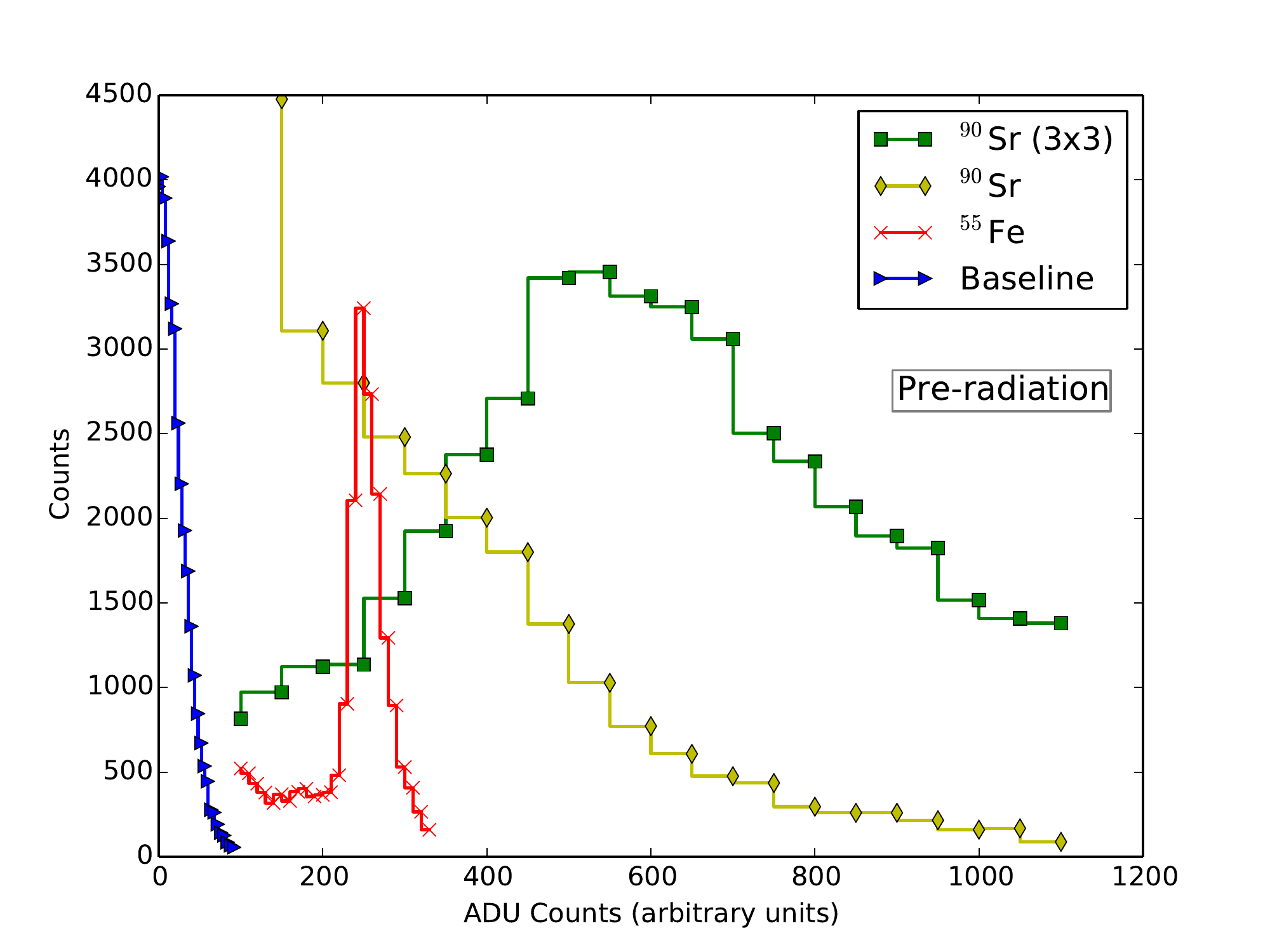}
\caption{Single pixel spectra for 25~$\mu$m pixel from $^{55}$Fe and $^{90}$Sr (single pixel and 3x3 clustered) radiative source at 150V bias and -20\celsius.}
\label{fig:pre_rad_Fe55}
\end{figure}

\begin{figure}
\centering
\includegraphics[width=0.5\textwidth]{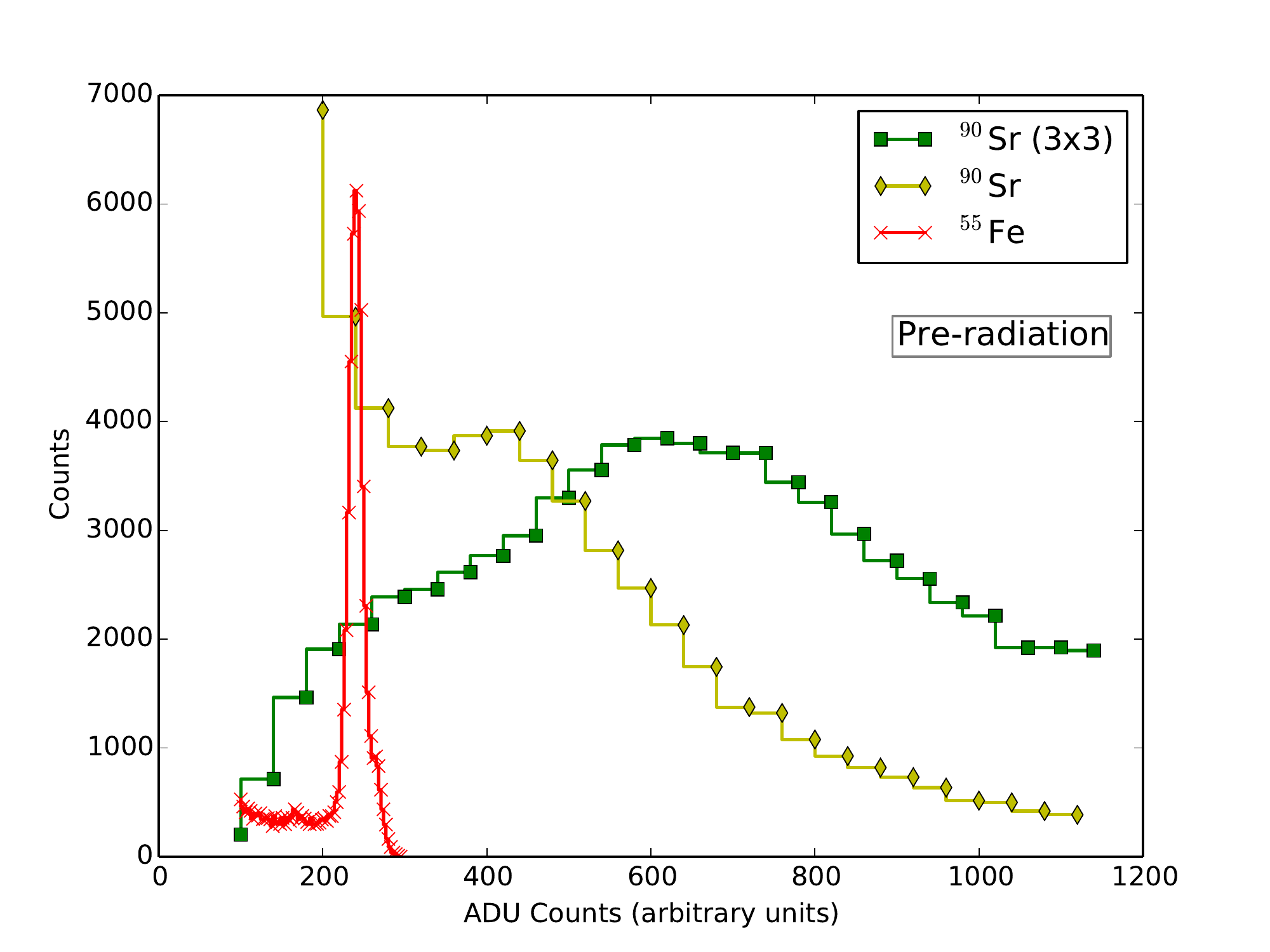}
\caption{Single pixel spectra for 50~$\mu$m pixel from $^{55}$Fe and $^{90}$Sr (single pixel and 3x3 clustered) radiative source at 150V bias and -20\celsius.} 
\label{fig:pre_rad_fe55_sr90}
\end{figure}

Spectrum measurements with radioactive sources of $^{55}$Fe and $^{90}$Sr have been conducted with pre-irradiated devices. Figure~\ref{fig:pre_rad_Fe55} and~\ref{fig:pre_rad_fe55_sr90} shows the result on 25 and 50~$\mu$m pitch pixel. Based on the 5.9~keV $^{55}$Fe calibration peak, the input capacitance has been estimated as $\sim$15~fF (gain of ~11$\mu$V/e). The noise was measured at about 30 electron equivalent noise charge (ENC) at baseline. We stress that the readout is not optimized for leakage current or input capacitance.  
From the $^{90}$Sr spectrum one can estimate for the Most Probable Value (MPV) a 3000-4000~e$^-$ signal which suggests a collection volume of about 40-50~$\mu$m at 150~V. For all plots a threshold of 100 ADUs (~700~e$^-$) is used for the cluster reconstruction (same in the seed and in the neighbor pixels).

\subsection{Sensor performance after neutron irrational}

The chips have been irradiated in the nuclear reactor at Ljubljana with 5 different doses of neutrons. The chips were not pre-characterized. Measurements for different doses have been conducted with different devices. The neutron and TID doses are shown in Table \ref{table:irrad}. The performance of the HV-SOI pixel sensor has been studied with radioactive sources of $^{90}$Sr and $^{55}$Fe.

\begin{figure}
\centering
\includegraphics[width=0.5\textwidth]{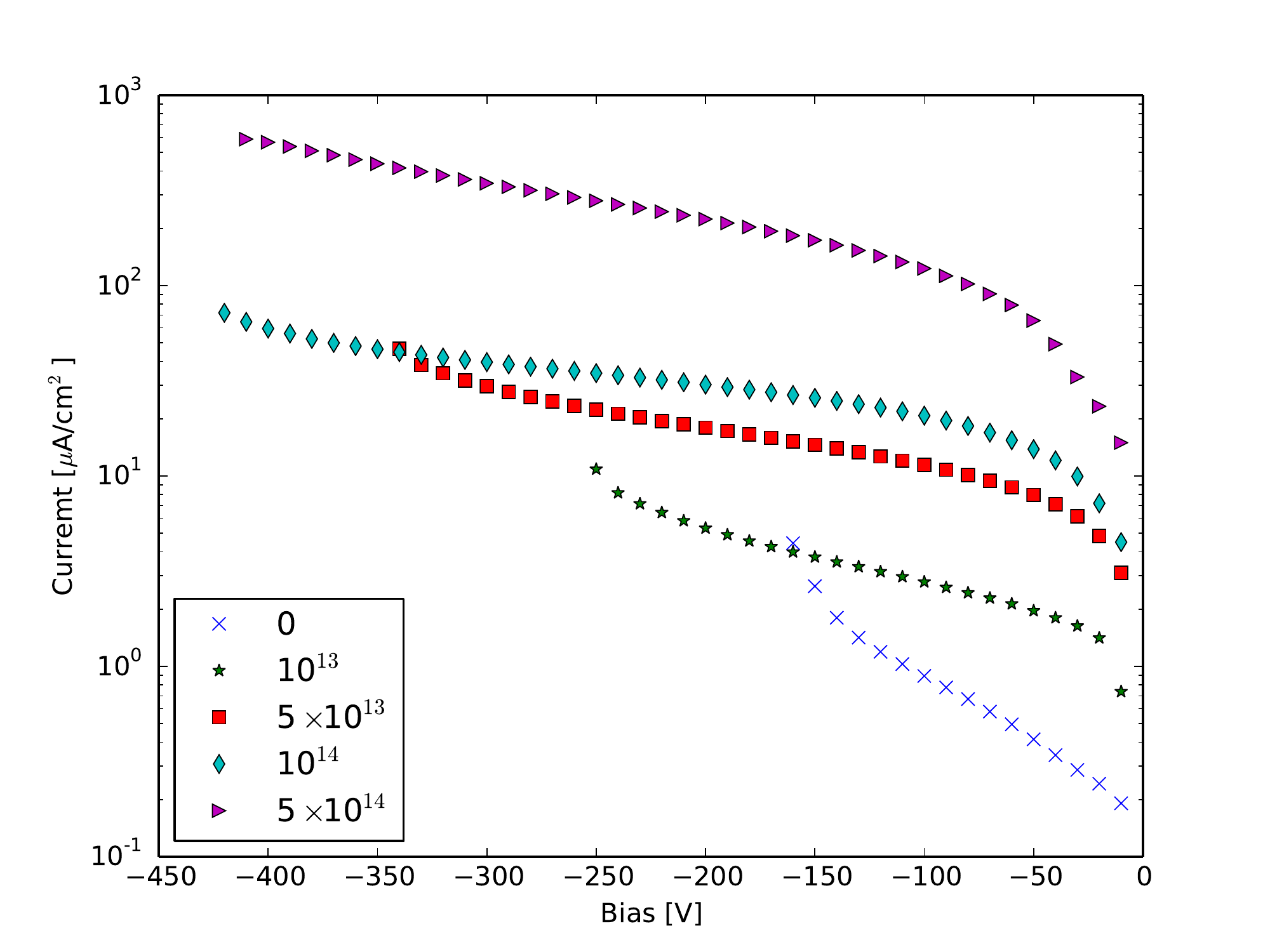}
\caption{I-V characteristics for XTB01 for different neutron dose at 25~C$^{\circ}$.}
\label{fig:iv}
\end{figure}

Measured I-V characteristics for an entire chip for different neutron doses at room temperature are depicted in Figure~\ref{fig:iv}. 
With a bias voltage of 150~V on 100 $\Omega\cdot$cm p-type substrate we expect about 50~$\mu$m depletion thickness. 
We can observe a linear increase of leakage current due to defects caused by neutron damage, while the breakdown voltages are increasing.
Defects could act as recombination/generation centers and are responsible for an increase of the leakage current. 

\begin{figure}
\centering
\begin{subfigure}{.5\textwidth}
  \centering
  \includegraphics[width=.95\linewidth]{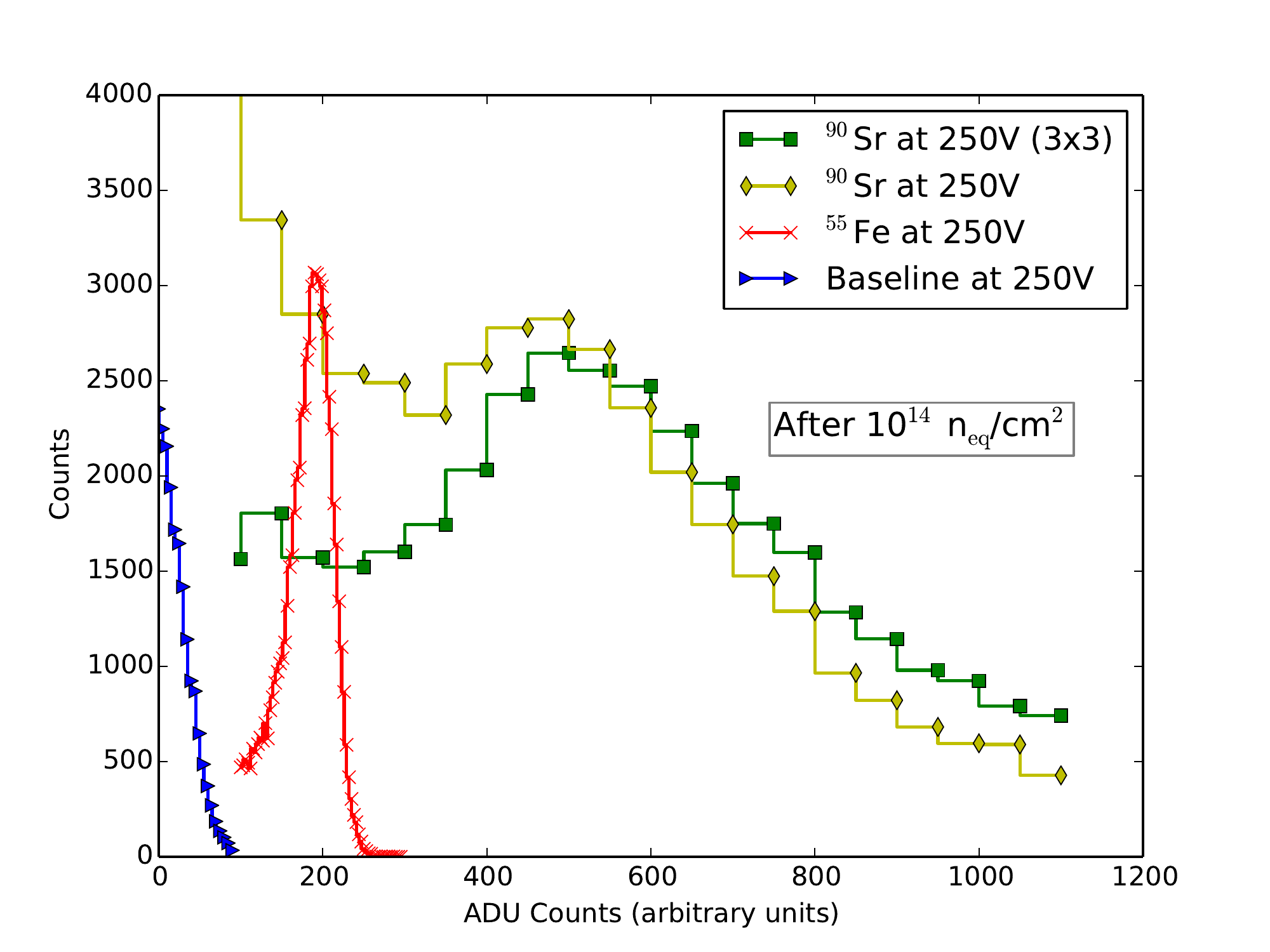}
  \caption{}
\end{subfigure}%

\begin{subfigure}{.5\textwidth}
  \centering
  \includegraphics[width=.95\linewidth]{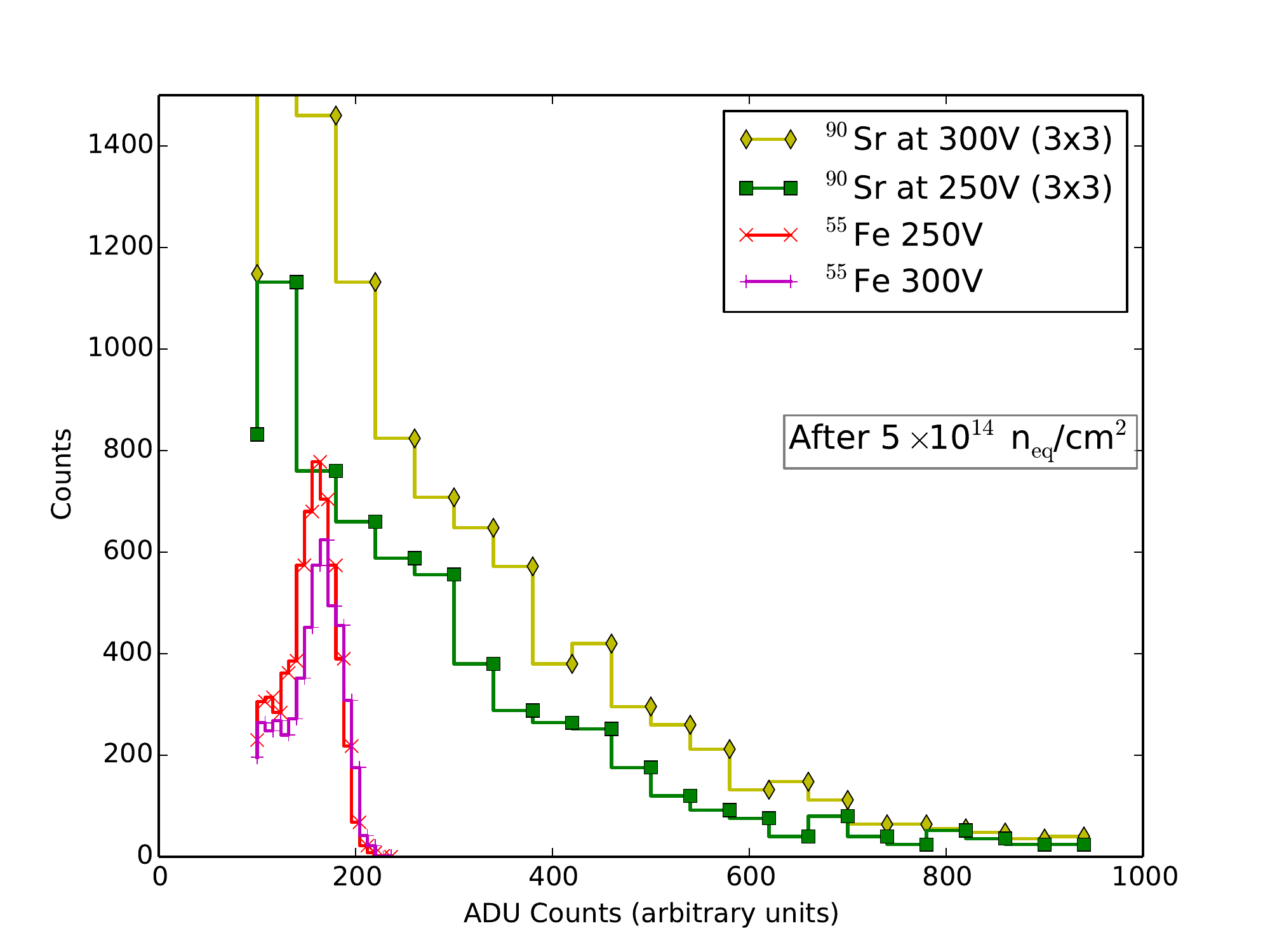}
  \caption{}
\end{subfigure}
\caption{Single pixel spectra for $^{55}$Fe and 3x3 clustered $^{90}$Sr for 50~$\mu$m pixel after a) 10$^{14}$ b) 5x10$^{14}$ n$_{\rm eq}$/cm$^{2}$}
\label{fig:rad_fe55_sr90_50}
\end{figure}

\begin{figure}[h!]
\centering
\begin{subfigure}{.5\textwidth}
	\centering
	\includegraphics[width=.95\textwidth]{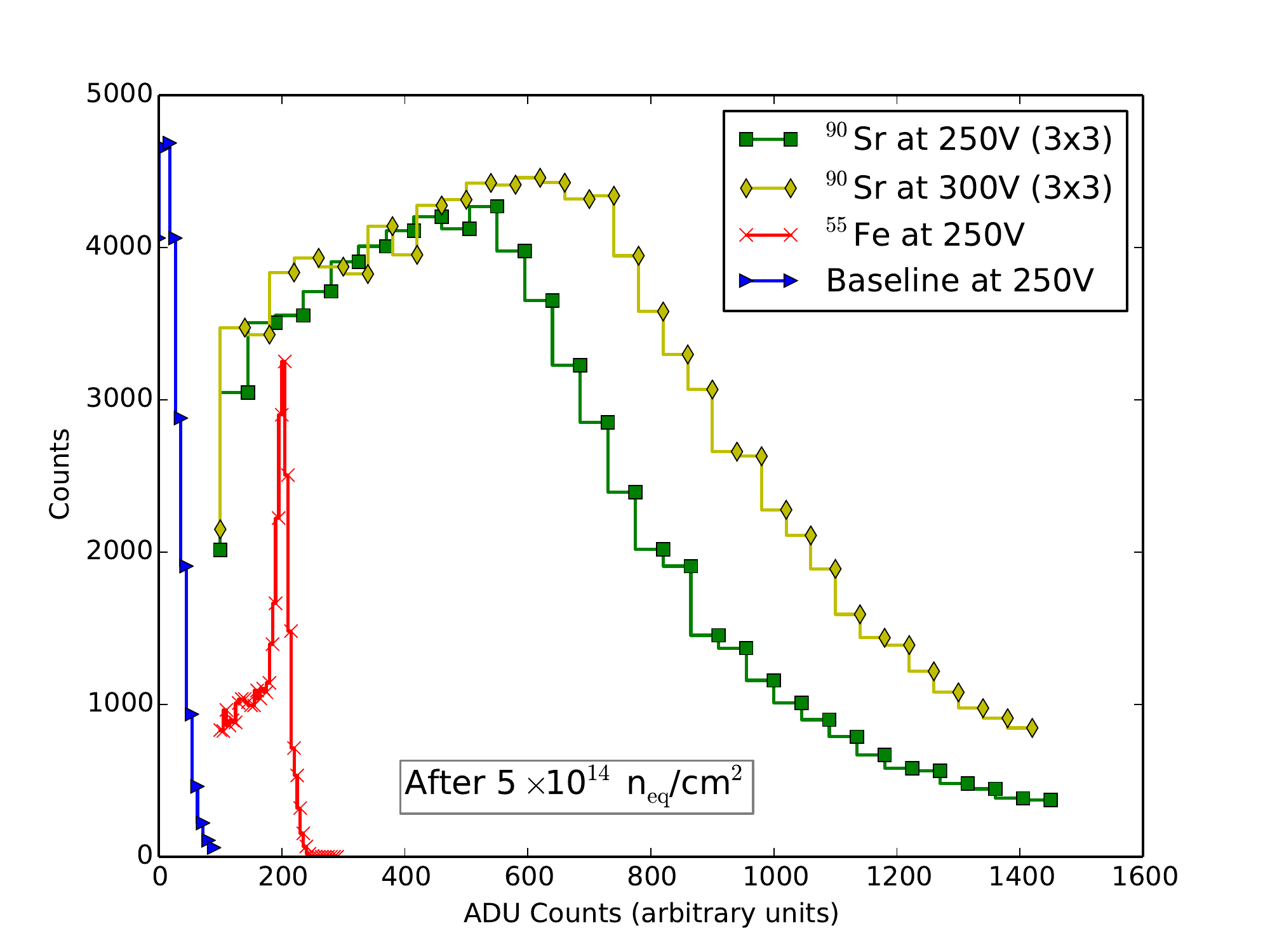}
	\caption{}
\end{subfigure}%

\begin{subfigure}{.5\textwidth}
  \centering
  \includegraphics[width=.95\linewidth]{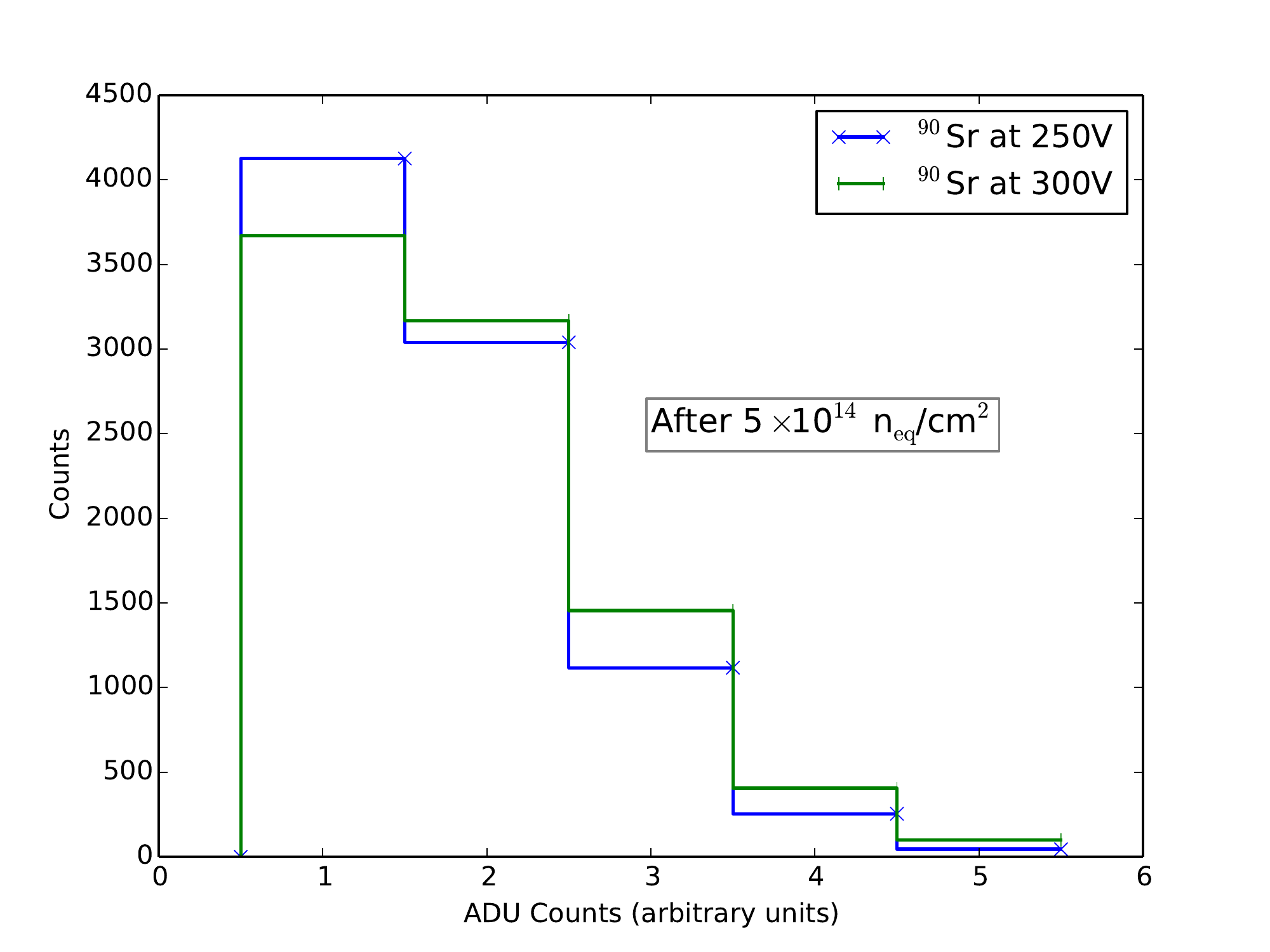}
  \caption{}
\end{subfigure}
\caption{a) 3x3 clustered pixel $^{90}$Sr and single cluster $^{55}$Fe and spectra for 25~$\mu$m pixel after 5x10$^{14}$ n$_{\rm eq}$/cm$^{2}$ and b)$^{90}$Sr cluster size distribution}
\label{fig:5e14_Sr90_25um}
\end{figure}

Figure~\ref{fig:rad_fe55_sr90_50} shows a $^{90}$Sr source spectrum after 10$^{14}$ and $5\times10^{14}$ n$_{\rm eq}$/cm$^{2}$ for 50~$\mu$m pixel. One can observe a clustered signal of about 4000~e$^-$ at 250~V after $10^{14}$ n$_{\rm eq}$/cm$^{2}$. No signal is observed for $5\times10^{14}$ n$_{\rm eq}$/cm$^{2}$ which may suggest inefficiencies between pixels caused by trapping. Losing charge due to trapping between pixels may be an effect of a large distance between collecting diodes and insufficient electrical field (see Figure~\ref{fig:pix_box_layout}). Figure~\ref{fig:5e14_Sr90_25um} shows $^{90}$Sr spectrum from clustered 25~$\mu$m pixels and cluster distribution for $5\times10^{14}$ n$_{\rm eq}$/cm$^{2}$. Signals of about 4000~e$^-$ are clearly seen, and this result would suggest to confirm that charge is not fully collected between pixels in case of 50~$\mu$m pixel pitch and $5\times10^{14}$ n$_{\rm eq}$/cm$^{2}$.

\section{Summary}

An improved SOI-MAPS for ionizing radiation based on HV-SOI technology has been developed. In comparison to existing SOI devices, this technology makes use of thick epitaxial layer and multi-well structures to isolate transistor channels from the BOX and make them immune to the Back Gate Effects. Access to a handling wafer below the BOX allows using the substrate as a particle sensing device. First measurements with a 100~$\Omega\cdot$cm handling wafer indicate that more than 200~V biased can be applied to the sensor. Signal from about 50~$\mu$m depleted part can be collected after $5\times10^{14}$ n$_{\rm eq}$/cm$^{2}$. Those first measurements indicate an encouraging prospect to use this technology for particle detection and tracking at radiation harsh environments. 
More detailed measurements are planned involving test-beam campaigns to investigate systematic studies in pixel efficiency. Comparison with TCAD simulations is also planned as a next step.
A second simple passive prototype chip has been submitted for detailed investigation on different guard ring structures and pixel diode geometries, especially focusing on isolation between pixels and technological changes.

\section*{Acknowledgments}

The authors acknowledge help provided by X-FAB especially K. Bach and A. H\"olke and a very fruitful collaboration with the CERN group especially S. Fernandez-Perez, M. Backhaus and H. Pernegger. Work supported by the German Research Foundation (DFG) under the  under contract no.~WE~976/4-1.

\section*{References}



\bibliographystyle{apsrev}

\end{document}